\documentstyle[aps,prl,multicol,epsf]{revtex}
\begin{document}
\title{Inverse cascade in Charney-Hasegawa-Mima turbulence}
\author{G. Boffetta$^{1,2}$ \and F. De~Lillo$^{1}$
         and S. Musacchio$^{1,2}$}
\address{
  $^1$ Dipartimento di Fisica Generale, Universit\`a di Torino -
           Via Pietro Giuria 1, 10125 Torino, Italy\\
  $^2$ INFM Sezione di Torino Universit\`a - 
           Corso Raffaello 30, 10125 Torino, Italy}

\maketitle

\begin{abstract}
The inverse energy cascade in Charney-Hasegawa-Mima turbulence
is investigated. Kolmogorov law for the third order
velocity structure function is shown to be independent on the
Rossby number, at variance with the energy spectrum, as shown
by high resolution direct numerical simulations.
In the asymptotic limit of strong rotation, coherent vortices 
are observed to form at a dynamical scale which slowly grows with time.
These vortices form an almost  quenched pattern and induce strong
deviation form Gaussianity in the velocity field.  
\end{abstract}

\begin{multicols}{2}
\noindent
%%%%%%%%%%%%%%%%%%%%%%%%%%%%%%%%%%%%%%%%%%%%%%%%%%%%%%%%%%%%%%    
%\section{Section title}

The existence of an inverse cascade is the most remarkable property
of two dimensional turbulence.
It was predicted by Kraichnan \cite{K67} for Navier--Stokes equation:
as a consequence of inviscid enstrophy conservation,
energy is forced to flow to large scales.
The inverse cascade can be sustained only in presence
of an external forcing injecting energy at a characteristic scale
into the system.
At scales larger than the forcing the turbulent flow is
essentially random with Gaussian velocity difference statistics
following Kolmogorov scaling \cite{BCV00}.
Thus, the presence of the inverse energy cascade prevents
the formation of the large scale coherent structures observed
in the case of decaying turbulence \cite{McW84}.
                                                                 
Large scale coherent structures in presence of the inverse cascade
have been observed only in presence of a characteristic scale
breaking scale invariance. A well known example is the so-called
Bose-Einstein condensation, when the energy accumulates at
the largest available scale \cite{SY93} forming vortices at
the system size.
 
Another example of vortex formation is the quasi-crystallization
phenomenon observed in Charney-Hasegawa-Mima (CHM) turbulence,
a paradigm for both geostrophic motion in planetary
atmospheres \cite{C71} and drift-wave turbulence in magnetically
confined plasma \cite{HM78}. For the stream function
$\psi({\bf x},t)$ CHM equation is written as:
\begin{equation}
{\partial \over \partial t} (\nabla^2 \psi - \lambda^2 \psi )
+ J(\nabla^2 \psi,\psi) = D + F
\label{eq:1}
\end{equation}
where $J$ denotes the Jacobian and $D$ and $F$ are damping and
forcing respectively.
In this case, vortices have been observed to form, in a
quasi-crystal structure, at the intrinsic scale
$1/\lambda$, corresponding to the Rossby deformation radius
in the atmosphere \cite{Salmon98} or to the effective ion Larmor
radius in plasma \cite{KOY95}.
 
In this Letter we focus on dynamics on scales much larger than 
$\lambda^{-1}$. In this regime there is no intrinsic scale involved in
the evolution of the system, which nevertheless exhibits formation of 
strong vortices. In this case we observe that the scale of vortices 
is a dynamical one which increases in time as a consequence of 
vortex merging.
The characteristic time of evolution slows down leading, in the
limit of large Reynolds numbers, to a disordered pattern
of quenched vortices.
Despite the presence of strong vortices, we find that the
two-dimensional $3/2$ Kolmogorov law for the third-order
velocity structure function holds, independently on the
value of $\lambda$. As a consequence, the kinetic energy spectrum
follows Kolmogorov scaling but with a different
constant with respect to the Navier--Stokes turbulence.   

The CHM equation (\ref{eq:1}) has two quadratic
inviscid invariants corresponding to total energy
\begin{equation}
E = E_{k}+ E_{\lambda} =
{1 \over 2} \langle (\nabla \psi)^2 + \lambda^2 \psi^2 \rangle
\label{eq:2}
\end{equation}
where $\langle ... \rangle$
denotes spatial average,
and total enstrophy
\begin{equation}
Z = Z_{k}+ Z_{\lambda} =
{1 \over 2} \langle (\nabla^2 \psi)^2 +
\lambda^2 (\nabla \psi)^2 \rangle
\label{eq:3}
\end{equation}
Both the inviscid invariants consist of two terms, the first corresponding
to kinetic contribution and the second to potential one.
The kinetic terms are, by definition, the only ones which survive in the
Navier--Stokes limit $\lambda \to 0$.
 
The range of scales are separated by the characteristic wavenumber
$\lambda$. For $k \gg \lambda$ the kinetic contributions dominate
in (\ref{eq:2}-\ref{eq:3}), at very large scales $k \ll \lambda$
the leading terms are the potential ones.
In the following we will assume that the forcing $F$ is
limited to a narrow band of wavenumber around $k_f$.
This will be the other relevant wavenumber in our problem.               

Kolmogorov-like dimensional analysis can be easily extended to the present
problem \cite{KOY95,LM91,OK92}. If $k_f \gg \lambda$ we
recover the well-known energy spectra for two-dimensional Navier--Stokes
turbulence with energy spectrum $E(k) \propto k^{-5/3}$ and
$E(k) \propto k^{-3}$ for $\lambda < k <k_f$ and $k > k_f$ respectively.
When $k_f \ll \lambda$ one obtains the prediction
$E(k) \propto k^{-11/3}$ for $k < k_f$ 
and $E(k) \propto k^{-5}$ for $k_f > k > \lambda$. 
Dimensionally predicted spectra have been confirmed by direct
numerical simulations \cite{KOY95,FM79,WFI97} but little is known
about structure functions and probability density functions.
 
The starting point for a statistical approach to turbulent cascade
is the energy flux, written in the physical space as \cite{F95}
\begin{equation}
\varepsilon(\ell) = - {\partial \over \partial t} {1 \over 2}
\langle {\bf \nabla} \psi({\bf x}) \cdot {\bf \nabla} \psi({\bf x}+\ell)
+ \lambda^2 \psi({\bf x}) \psi({\bf x}+\ell) \rangle|_{NL}
\label{eq:4}
\end{equation}
where the subscript $NL$ stands for the nonlinear contribution
of (\ref{eq:1}) to the time derivative.
Making use of (\ref{eq:1}) and of integration by parts we easily obtain
\begin{equation}
\varepsilon(\ell) = {1 \over 2}
\langle \psi({\bf x}) J(\psi({\bf x}+\ell),\nabla^2 \psi({\bf x}+\ell))
\rangle + ({\bf x} \leftrightarrow {\bf x}+\ell)
\label{eq:5}
\end{equation}
in which the parameter $\lambda$ has formally disappeared.
As a consequence we can make use of the well known result
for Navier--Stokes (i.e. $\lambda=0$) and write \cite{Y99}
\begin{equation}
\langle (\delta u_{L})^3 \rangle = {3 \over 2} \varepsilon \ell
\label{eq:6}                     
\end{equation}
where $\delta u_{L}$ represents the longitudinal increment of
the velocity ${\bf v}=(\partial_y \psi,-\partial_x \psi)$
and $\varepsilon$ is the energy input due to the forcing.
We thus have a new degree of universality for the 3/2 Kolmogorov
law in two-dimensional turbulence, with respect to the class
of equations (\ref{eq:1}) parameterized by $\lambda$.
From a dimensional point of view, (\ref{eq:6}) implies a
scaling exponent $h=1/3$ for velocity increments and
thus a scaling exponent $4/3$ for $\psi$, as in Navier--Stokes turbulence.
From (\ref{eq:2})
one obtains the different predictions for the spectrum
discussed above. In particular, the kinetic spectrum has
the form
\begin{equation}
E_{k}(k) = C_{\lambda} \varepsilon^{2/3} k^{-5/3}
\label{eq:7}
\end{equation}    
for any value of $\lambda$, but with a Kolmogorov constant $C_{\lambda}$
which, in principle, can depend on $\lambda$.
A simple physical argument for this dependency is as follows.
The scaling of the eddy turnover time depends on the scaling exponent.
In the kinetic limit $k \gg \lambda$ one has the standard Kolmogorov
scaling $\tau(k) \simeq \varepsilon^{-1/3} k^{-2/3}$ \cite{F95}.
On the other hand,
in the potential limit $k \ll \lambda$, (\ref{eq:1}) gives
$\tau(k) \simeq (\lambda/k)^2 \varepsilon^{-1/3} k^{-2/3}$.
Thus, for $k \ll \lambda$, the efficence of energy transfer is reduced
and one expects a larger value of the constant in (\ref{eq:7}).

We have numerically investigated the inverse cascade in the potential
energy regime by direct numerical simulations.
In order to avoid complications induced
by the crossover from the kinetic domain to the potential domain,
we study the system in the limit $\lambda \to \infty$. This is to be
seen only as a formal procedure,  equivalent to considering
wavenumbers much smaller than $\lambda$, which physically might be the 
case for magnetized plasma in presence of a strong magnetic field.
Indeed this limit provides us with a model suitable for any $\lambda 
\gg k_f$: because the energy is transferred to large scale, the 
dominance of the potential term will be assured in all the inertial 
range.  In the limit $\lambda \gg k_f$, rescaling the time $t \to 
t/\lambda^2$, one obtains the so-called asymptotic model \cite{LM91}
\begin{equation}
{\partial \psi \over \partial t} = J(\nabla^2 \psi,\psi) + D + F
\label{eq:8}
\end{equation}
for which the conserved quantities become
\begin{eqnarray}
E &=& {1 \over 2} \langle \psi^2 \rangle \nonumber \\    
& & \\
Z &=& {1 \over 2} \langle (\nabla \psi)^2 \nonumber \rangle
\label{eq:9}
\end{eqnarray}
\begin{figure}
\epsfxsize=8truecm
\epsfbox{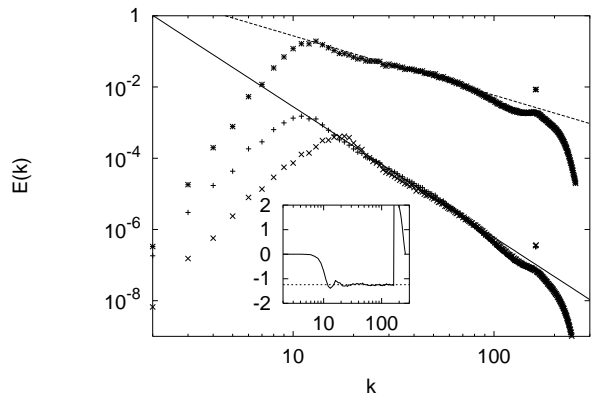}
\narrowtext
\caption{Potential energy spectrum $E_{\lambda}(k)$ at times
$t=4 \times 10^{-3}$ ($\times$) and $t=10^{-2}$ ($+$) and
kinetic energy spectrum $E_{k}(k)$ at time $t=10^{-2}$
($*$) averaged over $14$ realizations.
The continuous line represents the dimensional
prediction $k^{-11/3}$, the dashed line is the Kolmogorov spectrum
$E_{k}(k)= C_{\lambda} \varepsilon^{2/3} k^{-5/3}$ with constant
$C_{\lambda}=11$.
In the inset we plot the average energy flux at time $t=10^{-2}$
with the line $\epsilon=1.24$.}
\label{fig1}
\end{figure}

We have integrated (\ref{eq:8}) with a standard pseudo-spectral code
in a double periodic domain of size $L=2 \pi$ at resolution
$N=512$. The forcing is white in time in a narrow band of wavenumbers
around $k_f=160$.
The dissipative term in (\ref{eq:8}) has the role of removing
potential enstrophy at small scales and, as customary, it is numerically
substituted by a hyperviscous term (of order $8$ in our
simulations).  Time evolution is obtained by a standard second-order
Runge-Kutta scheme starting from a zero initial condition. The
run is stopped at a given time $T$ at which the energy containing
scales are still much smaller than the computational box in order
to avoid condensation effects \cite{SY93} (see Figure~\ref{fig1}).   
All the results discussed in the following are taken after
averaging over $14$ independent realizations.
 
The limitation in the resolution ($N=512$)
is due to the discussed scaling of the characteristic time.
Even with this moderate resolution, the ratio of the large scale
characteristic time with the forcing scale time is about $2000$
and thus time evolution is very expensive ($10^6$ time
steps for each realization). In the case of Navier--Stokes
turbulence ($\lambda=0$) this would correspond to an integration 
covering about $5$ decades of inertial range.
 
In Figure~\ref{fig1} we plot the potential energy spectrum
$E_{\lambda}(k)$ at two different times.
The scaling exponent $-11/3$ is clearly
visible even if some accumulation at the largest mode is
evident. This accumulation is not due to condensation as it is still
well below the largest mode and it moves in time.
We think that the existence of this ``bump'' is a genuine effect,
probably due to the rapid growth of characteristic times and
to the presence of intense vortices, as discussed below.
The energy flux $\varepsilon \simeq 1.24$ is estimated by
the plateau of the energy flux shown in the inset.                     

\begin{figure}
\epsfxsize=8truecm
\epsfbox{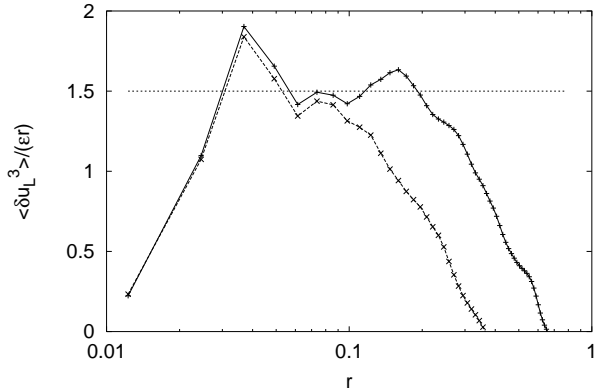}
\caption{Third-order longitudinal structure function
$\langle (\delta u_{L}(r))^{3} \rangle$ compensated with
the dimensional prediction $\varepsilon r$ at times
$t=4 \times 10^{-3}$ ($\times$) and $t=10^{-2}$ ($+$) averaged
over $14$ independent realizations.
The line represents the Kolmogorov law (\ref{eq:6}).}
\label{fig2}
\end{figure} 

The ``$3/2$'' law for the longitudinal velocity structure function
is shown in Figure~\ref{fig2}, also plotted at two
different times.
The compensation with the theoretical prediction (\ref{eq:6})
is remarkable, taking into account the limited
resolution of our runs. As expected, the extension of the inertial range
increases with time without changing small scale statistics.
The oscillations observed at small scales are due to the
contamination of the forcing. A similar effect was observed
also in NS simulations.

As discussed above, the fact that the ``$3/2$'' law is independent
on $\lambda$ (and thus the velocity scaling exponent has
always the Kolmogorov value $h=1/3$) do not imply that the statistics,
and in particular the form of the pdf of velocity differences,
is the same as for Navier--Stokes equation.
For example, in Figure~\ref{fig1} we also plot the kinetic energy spectrum
$E_{k}(k)$ at final time $t=10^{-2}$.
The scaling exponent is compatible with the Kolmogorov value
$5/3$ as predicted by (\ref{eq:7}),
but the Kolmogorov constant $C_{\lambda} \simeq 11$ is about two times
that of Navier--Stokes \cite{BCV00}. A larger constant means
a suppression of the energy flux which is a direct consequence
of the dilatation of the characteristic times. 
                 
\begin{figure}
\epsfxsize=8truecm
\epsfbox{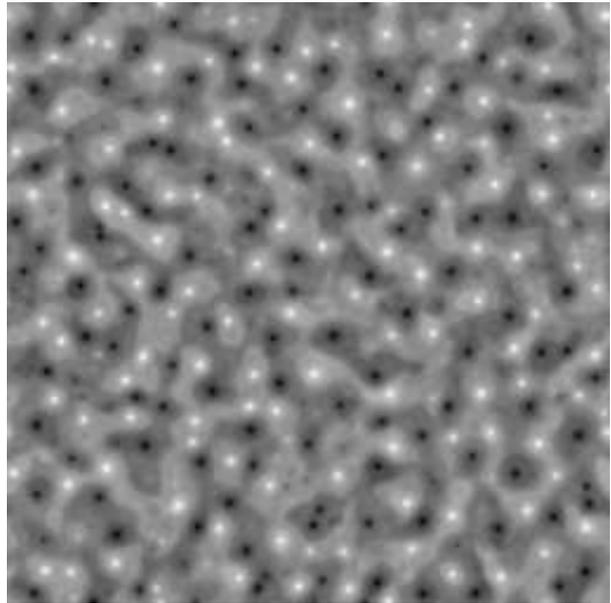}
\caption{Grayscale plot of the stream function $\psi$
at time $t=10^{-2}$. The characteristic scale of vortices
$\ell_E \sim L/10$ corresponds to the peak of the energy
spectrum in Figure~\ref{fig1}.}
\label{fig3}
\end{figure}
 
A more significant difference with respect to Navier--Stokes inverse
cascade is the presence of strong vortices, as shown in Figure~\ref{fig3}.
Vortices in CHM turbulence are injected at the forcing scale and they
organize themselves to form a random pattern on the
characteristic scale $\ell_{E}$.
This dynamical scale is associated to the peak of the spectrum of
Figure~\ref{fig1}. Vortex dynamics slows down as $\ell_{E}$
increases as $\tau(\ell_E) \sim \ell_E^{8/3}$. Thus, in the
limit of large Reynolds number the system will end in a disordered
pattern of quenched vortices forming a kind of ``turbulent glass''.   

\begin{figure}[p]
\epsfxsize=8truecm
\epsfbox{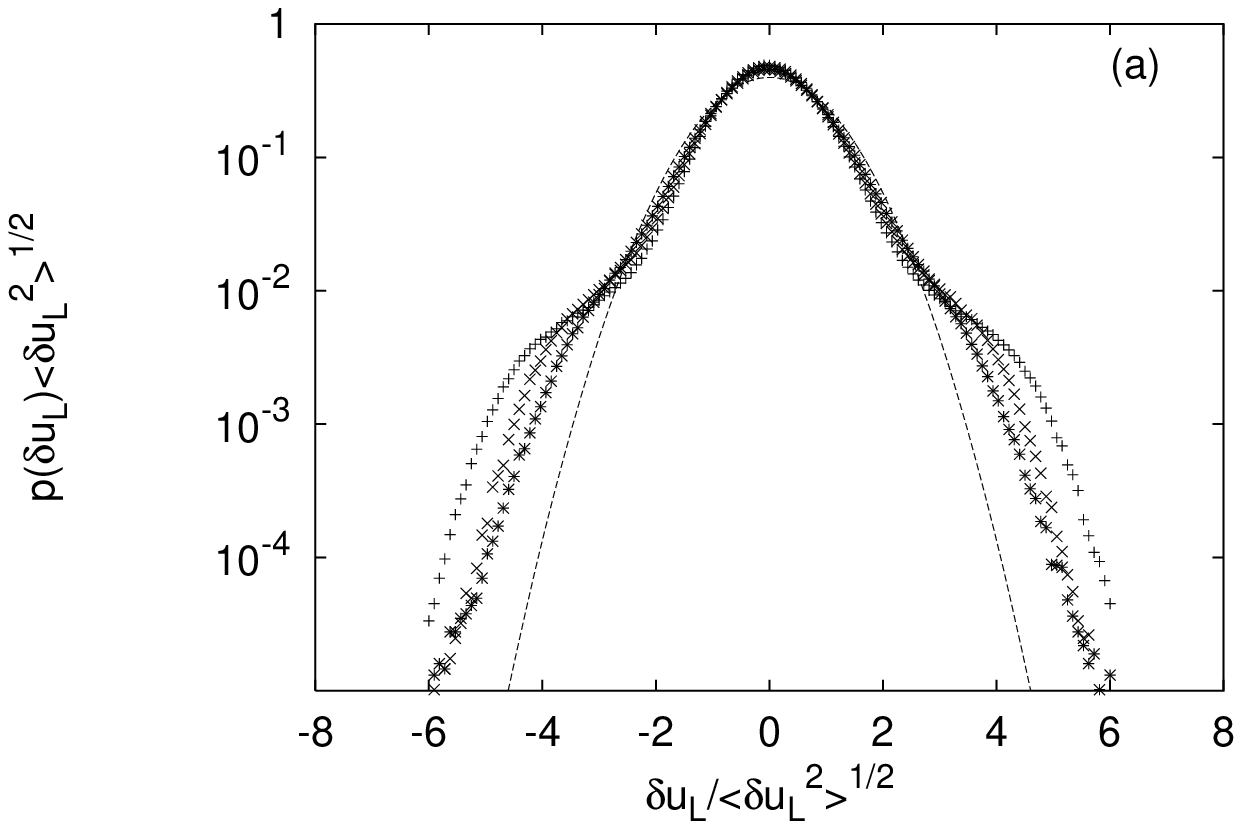}
\epsfxsize=8truecm
\epsfbox{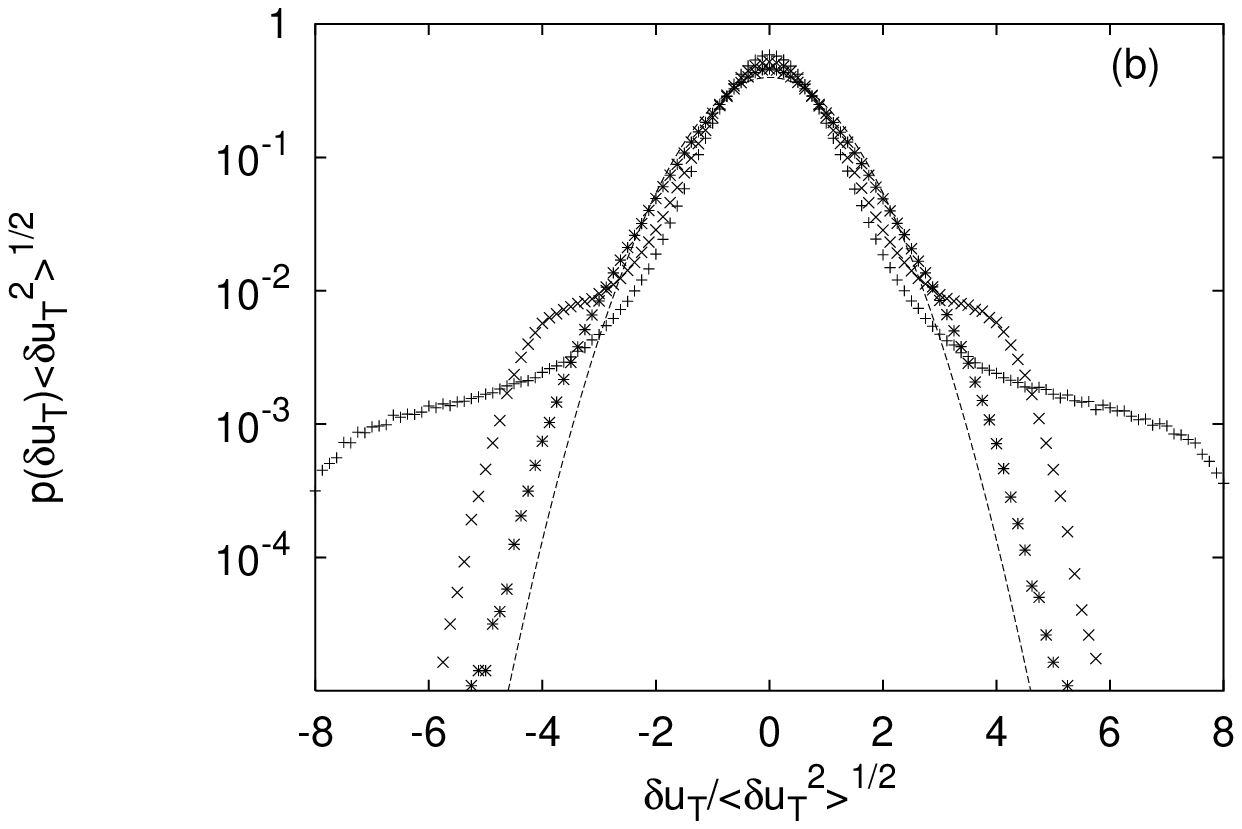}
\caption{Probability density functions of longitudinal (a)
and transverse (b) velocity differences at separations
$\ell=0.05$ ($+$), $\ell=0.1$ ($\times$) and $\ell=0.2$ ($*$)
at time $t=10^{-2}$. Dashed line represents Gaussian distribution.}
\label{fig4}
\end{figure}

An important consequence of the presence of strong vortices
is that the statistics of the velocity field strongly deviates from
Gaussianity. In Figure~\ref{fig4} we plot the pdf of longitudinal and
transverse velocity differences at three different scales within
the inertial range. The effect of vortices is evident by the
presence of large ``wings'' in the tails, in particular
on the transverse velocity differences which are more sensible
to a rotating structure.
 
In conclusion, we have shown that Kolmogorov ``3/2'' law for
two-dimensional energy cascade in Charney-Hasegawa-Mima turbulence is
independent on the value of the intrinsic scale $\lambda$.
Velocity statistics satisfies Kolmogorov scaling with non-universal
coefficients. In the asymptotic limit $\lambda \to \infty$
the Kolmogorov constant is found to be about $2$ times the
Navier--Stokes case \cite{OK92}.
Strong coherent vortices are found to emerge at the forcing scale
and aggregate to form a pattern of quenched vortices at large scale.
As a consequence of the presence of vortices, strong deviations
from Gaussianity are observed in the velocity field.
 
%%%%%%%%%%%%%%%%%%%%%%%%%%%%%%%%%%%%%%%%%%%%%%%%%%%%%%%%%%%%%%
\acknowledgments
This work was partially supported by MURST
Cofin 2001 n.2001023848.
We acknowledge the allocation of computer resources
from INFM Progetto Calcolo Parallelo.
%%%%%%%%%%%%%%%%%%%%%%%%%%%%%%%%%%%%%%%%%%%%%%%%%%%%%%%%%%%%%%

\end{multicols}
\end{document}